\documentclass{article}
\usepackage{amsmath,amssymb,graphicx}
\graphicspath{{../figs/}}
\usepackage{epstopdf}
\usepackage[colorlinks=true]{hyperref}
\usepackage[margin=1in]{geometry}
\usepackage{mathtools}
\usepackage{dsfont}
\usepackage{bm}
\usepackage{nccmath,xfrac}
\usepackage{lineno}
\usepackage[draft]{changes}
\usepackage{enumitem}
\usepackage{authblk}
\usepackage{comment}
\usepackage{xcolor}
\usepackage{float}
\newcommand{\n}{\bm{n}}
\newcommand{\dd}{\mathrm{d}}

\title{A generalized work theorem for stopped stochastic
  chemical reaction networks}

\author[1]{Xiangting Li}
\author[1,2]{Tom Chou} \affil[1]{Department of Biomathematics,
  University of California, Los Angeles, CA 90095}
\affil[2]{Department of Mathematics, University of California, Los
  Angeles, CA 90095}
\date{\today}
\begin{document}
\maketitle


\begin{abstract}
  We establish a generalized work theorem for stochastic chemical
  reaction networks (CRNs). By using a compensated Poisson jump
  process, we identify a martingale structure in a generalized entropy
  defined relative to an auxiliary backward process and
  extend nonequilibrium work relations to processes stopped at
  bounded arbitrary times.  Our results apply to discrete, mesoscopic chemical
  reaction networks and remain valid for singular initial conditions
  and state-dependent termination events.  We show how martingale
  properties emerge directly from the structure of reaction
  propensities without assuming detailed balance. Stochastic
  simulations of a simple chemical kinetic proofreading network are
  used to explore the dependence of the exponentiated entropy
  production on initial conditions and model parameters, validating our
  new work theorem relationships. Our results provide new quantitative
  tools for analyzing biological circuits ranging from metabolic to
  gene regulation pathways.
\end{abstract}


\section{Introduction}
The Jarzynski equality,

\begin{align}
\mathds{E}\big[e^{-\beta W}\big] = e^{-\beta\Delta F},
\label{eq:jarzynski}
\end{align}
also known as the work theorem \cite{Jarzynski1997April}, and the
Crooks fluctuation theorem \cite{Crooks1999September} revolutionized
our understanding of non-equilibrium thermodynamics by relating
irreversible work to equilibrium free energy differences. While these
results were initially formulated for macroscopic systems, recent
extensions to mesoscopic scales \cite{Schmiedl2007January}, in
particular chemical reaction networks, have enabled possible
applications to biological systems.

The work theorem becomes particularly relevant in the context of
recognition and proofreading, where the energy of binding and the
information content of a signal are intimately connected. It is known
that in the setting of kinetic proofreading, energy is dissipated to
enhance the fidelity of ligand-receptor recognition. Although a
competition between energy dissipation and signal fidelity is known in
specific networks, a universal tradeoff across
different networks exists remains an open question. Work theorems or
stochastic thermodynamics in general could possibly provide a
framework to address this question.


Nonetheless, work theorems such as Jarzynski's equality for chemical
reaction networks only apply to entropy production between two
specified time points \cite{Esposito_crn_2016,Esposito2018}. This
restriction hinders the application of work theorems to processes such
as kinetic proofreading since the process stops once a specified final
state is reached, not after a specified time has elapsed.

Martingales are a key tool that can be used to derive identities for
stochastic quantities after a process is stopped at different times
\cite{Neri2017,Roldn2023April,JSP2025}.  In our prior work
\cite{JSP2025}, we generalized the work theorem for overdamped and
underdamped Langevin dynamics to a family of martingales that hold
simultaneously for all bounded stopping times. In this paper, we
extend this framework to stochastic chemical reaction networks. This
extension is nontrivial because prior stochastic calculus tools are
not readily applicable to discrete systems \cite{Neri2017}.

We first introduce a compensated Poisson process to describe chemical
reaction networks that are a simplification of those treated by
Schmiedl and Seifert \cite{Schmiedl2007January}. We remove
unnecessary microscopic molecular details but retain the general
physical description of the chemical species.  Based on the
compensated Poisson process description, we can similarly define a
backward process $\psi(\mathbf{n},t)$, which satisfies the
time-reversed chemical master equation.

By using $\psi(\mathbf{n},t)$ in the place of probability density
$p(\mathbf{n},t)$ in the forward process, we can compute the time
derivative of the exponentiated negative entropy production
$\mathrm{d}\sigma(t)$ and split it into a deterministic part
($\mathrm{d}t$ terms) and a discrete martingale part
($\mathrm{d}\tilde{N}_i(t)$ terms). In the case of single-molecule
CRNs, where the numbers of stochastic species are either 0 or 1, the
deterministic part vanishes, suggesting a martingale structure.
Interestingly, for multi-molecule CRNs, the deterministic part does
not vanish, which sets the CRN framework apart from the previous work
on Langevin dynamics.

\section{System Setup}
We consider a reaction network in which a small number of
species—often only a single enzyme or receptor—undergo stochastic
transitions, while other species are ``chemostated'' and held at fixed
concentrations by large reservoirs.  Denote by $n_j(t)\in\{0,1\}$ the
random copy number of each \textbf{stochastic species} $j$, and by
$c_\alpha$ the constant concentration of each \textbf{chemostat
  species} $\alpha$. The chemostat reservoirs are characterized by a
uniform inverse temperature~$\beta$, so that every reaction must
respect microscopic detailed balance at that temperature.

Each elementary reaction channel is labeled by $i$, with a
corresponding reverse channel denoted $-i$. For reasons made clear
later, we use $N_i(t)$ to record the number of times reaction $i$ has
fired up to time $t$.  Associated with each firing of the $i^{\rm th}$
channel, the state vector $\mathbf{n}(t)$ jumps by $\Delta_i$. The
$j^{\rm th}$ component of the chemical state changes according to

\begin{equation}
  \label{eq:state_change}
  n_{j}(t) = n_{j}(0) + \sum_{i} \Delta_{i}^{j}N_{i}(t),
\end{equation}
where $\Delta_i^j=-\Delta_{-i}^j$ is the number of molecules of
species $j$ consumed or produced by reaction $i$, \textit{i.e.}, the
stoichiometric coefficient.
%
%
%
In cases where at most one copy of each stochastic species can exist,
any transition that would take $n_j$ outside $\{0,1\}$ is forbidden.

\subsection{Reaction Propensities}
The propensity of channel $i$ combines a thermodynamic factor and, in
general, a combinatorial factor. We associate an activation free
energy level $G_i^\ddagger$ for each channel $i$ (with
$G_{-i}^\ddagger=G_i^\ddagger$), assign each reaction species $j$ an energy
$E_j$, and each chemostat species $\alpha$ a chemical potential
\cite{Esposito_crn_2016}.

\begin{equation}
  \label{eq:chem_pot}
\mu_\alpha = \mu_{\alpha,0} + \mfrac{1}{\beta}\ln c_\alpha,
\end{equation}
where $c_\alpha$ is measured relative to a reference concentration
associated with the reference value $\mu_{\alpha,0}$. A general
expression for the propensity of reaction $i$, applicable to
multi-molecule systems, is given by

\begin{equation}
  \label{eq:propensity}
  \lambda_i(\mathbf{n}) = A_i e^{-\beta\Big(G_i^\ddagger -
    \!\!\!\sum\limits_{j:\Delta_i^j<0}\!\! E_j|\Delta_i^j| -
    \!\!\!\sum\limits_{\alpha:\Delta_i^\alpha<0}\!\!
    \mu_{\alpha,0}|\Delta_i^\alpha|\Big)} 
  \Bigg(\prod_{j:\Delta_i^j<0} \frac{n_j!}{(n_j-|\Delta_i^j|)!}\Bigg)
  \Bigg(\prod_{\alpha:\Delta_i^\alpha<0}
  c_\alpha^{|\Delta_i^\alpha|}\Bigg).
\end{equation}
Here, $A_i$ is a kinetic prefactor, assumed to be symmetric ($A_{-i} =
A_i$). The sums are taken over reactant stochastic species $j$ (for
which $\Delta_i^j<0$) and reactant chemostat species $\alpha$ (for
which $\Delta_i^\alpha<0$). The term
$\frac{n_j!}{(n_j-|\Delta_i^j|)!}$ represents the number of ways to
choose the $|\Delta_i^j|$ reactant molecules of species $j$ from the
$n_j$ available.  In a single-molecule system ($n_j \in \{0,1\}$), the
combinatorial factors $\frac{n_j!}{(n_j-|\Delta_i^j|)!}$ (where
$|\Delta_i^j|=1$ if $j$ is a reactant) simplify to $n_j$ or
equivalently a Kronecker delta $\delta_{n_j,1}$ factor that ensures
the reactant is present. The propensity then simplifies to

\begin{equation}
  \label{eq:propensity_single}
  \lambda_i(\mathbf{n}) = A_i e^{-\beta\Big(G_i^\ddagger -
    \sum\limits_{j:\Delta_i^j<0} E_j|\Delta_i^j| -
    \sum\limits_{\alpha:\Delta_i^\alpha<0}
    \mu_{\alpha}|\Delta_i^\alpha|\Big)}
  \prod_{j:\Delta_i^j<0}\delta_{n_j}^1.
\end{equation}
The propensity depends on the concentrations $c_\alpha$ of the
chemostated reactants through the chemical potential $\mu_{\alpha}$
defined in Eq.~\eqref{eq:chem_pot}.

%
%
%
%
%
%

\subsection{Jump Dynamics via Compensated Poisson Processes}

For each reaction $i$, we introduce a time-changed Poisson counting
process $N_i(t)$ with a state-dependent rate
$\lambda_i(\mathbf{n}(t))$.  The corresponding compensated process

\begin{equation}
  \label{eq:compensated}
  \widetilde{N}_i(t)=N_i(t)-\int_0^t\lambda_i
  \bigl(\mathbf{n}(s^{-})\bigr)\,\mathrm{d}s
\end{equation}
was previously introduced in \cite{Anderson201501} as a special case
of Theorem 3.11 in \cite{Kurtz1980August}. Here $\mathbf{n}(s^{-})$
denotes the state just before a jump at time $s$.  Using the
compensated Poisson process $\widetilde{N}_i(t)$, the stochastic state
vector $\mathbf{n}(t)$ evolves according to the jump SDE

\begin{equation}
  \label{eq:jump_SDE}
{n}_j(t)=n_j(0)+\sum_i\Delta_i^j\,N_i(t)
=n_j(0)+\sum_i\Delta_i^j\int_0^t
\lambda_i\big(\mathbf{n}(s^{-})\big)\,\mathrm{d}s
+\sum_i\Delta_i^j\,\widetilde{N}_i(t)\,,
\end{equation}
or, equivalently,

\begin{equation}
  \label{eq:jump_diff}
  \mathrm{d}n_j(t)=\sum_i\Delta_i^j
  \Bigl[\lambda_i\big(\mathbf{n}(t^{-})\big)\,\mathrm{d}t
    + \mathrm{d}\widetilde{N}_i(t)\Bigr].
\end{equation}

The compensated counting process $\widetilde{N}_{i}(t)$ in
Eq.~\ref{eq:compensated} isolates purely stochastic fluctuations of
reaction events by subtracting the expected number of firings up to
time $t$, conditioned on the past trajectory. By construction,
$\widetilde{N}_{i}(t)$ has zero drift and is a martingale. The jump
dynamics of the chemical reaction network can then be written as a sum
of deterministic drift terms and martingale noise terms, which will
facilitate the isolation of martingale structures in entropy
production and underlies the generalized work theorems derived below,
particularly for processes stopped at random times.

\subsection{Energy, Work, and Heat}
The internal energy $E(\mathbf{n})$ is defined by

\begin{equation}
  \label{eq:energy_def}
  E(\mathbf{n})=\sum_j E_j\,n_j,\quad
  \mathrm{d}E=\sum_{i,j}E_j\Delta_i^j\,\mathrm{d}N_i,
\end{equation}
while the work done on the system by the reservoirs is given by the
replenishment of the chemostat species to compensate for the molecules
consumed by the stochastic reactions.  Thus, using

\begin{equation}
  \label{eq:work_def}
\mathrm{d}W=-\sum_{i,\alpha}\mu_{\alpha}\Delta_i^\alpha\,\mathrm{d}N_i.
\end{equation}
in the first law of thermodynamics, $\mathrm{d}E=\mathrm{d}W-q$,  we find

\begin{equation}
  \label{eq:heat_def}
  q=-\mathrm{d}E+\mathrm{d}W
  =-\sum_{i,j}E_j\Delta_i^j\,\mathrm{d}N_i
  -\sum_{i,\alpha}\mu_{\alpha}\Delta_i^\alpha\,\mathrm{d}N_i,
\end{equation}
where $q$ denotes the time derivative of the total heat dissipated 
from the system to the reservoirs $Q(t) = \int_0^t q(s)$.

\subsection{Chemical Master Equation and Generalized Backward Entropy}
The probability $p(\mathbf{n},t)$ evolves according to the chemical master
equation \cite{Gillespie1992}

\begin{equation}
  \label{eq:master_eq}
  \partial_t p(\mathbf{n},t)=\sum_i
  \Big[\lambda_i(\mathbf{n}-\Delta_i)p(\mathbf{n}-\Delta_i,t)
    -\lambda_i(\mathbf{n})p(\mathbf{n},t)\Big].
\end{equation}
The stochastic information entropy can then be defined as
%

\begin{equation}\label{eq:info_entropy}
S(t)=-\ln p\big(\mathbf{n}(t),t\big),
\end{equation}
where we have adopted natural units in which $k_{\rm B}=1$ and
temperature is measured in units of energy.

Next, we introduce an auxiliary backward process, $\psi(\mathbf{n},t)$,
which evolves according to the ``backward'' master equation

\begin{equation}
\label{eq:backward_master}
-\partial_t\psi(\mathbf{n},t) = \sum_i
\Big[\lambda_i(\mathbf{n}-\Delta_i)\psi(\mathbf{n}-\Delta_i,t)
  - \lambda_i(\mathbf{n})\psi(\mathbf{n},t)\Big].
\end{equation}
Analogous to $S(t)$, we define a generalized entropy $\Sigma(t)$
associated with this auxiliary process:

\begin{equation}
  \label{eq:gen_entropy}
  \Sigma(t) \coloneqq -\ln\psi\big(\mathbf{n}(t),t\big).
  \end{equation}
%

Entropy production refers to the part of the entropy difference
$\mathrm{d} S(t)$ that is not due to heat exchange $-\beta q(t)$ and
is evaluated by $\mathrm{d}S(t) + \beta q(t)$. For the generalized
entropy $\Sigma(t)$, we define the generalized entropy production
$\theta(t)$ by the differential equation

\begin{equation}
  \label{eq:gen_prod}
\mathrm{d}\theta(t) = \mathrm{d}\Sigma(t) + \beta q(t)\,.
\end{equation}
Using It\^o's formula on $\Sigma(t)$ for the jump process, we find

%
%
\begin{equation}
\dd \Sigma(t) = \frac{\partial \Sigma(t)}{\partial t}\,\dd t 
  -\sum_{i} \Big[\ln \psi(\n+\Delta_{i},t) -\ln\psi(\n,t)\Big]\dd N_{i}.
\end{equation}
After applying Eq.~\ref{eq:backward_master} and using Eq.~\ref{eq:heat_def}
for the dissipated heat $q$, we have
\begin{equation}
\begin{aligned}
  \mathrm{d}\theta(t) = &
  \sum_i \bigg[\frac{\psi\bigl(\mathbf{n}(t^{-})-\Delta_i,t\bigr)}
    {\psi\bigl(\mathbf{n}(t^{-}),t\bigr)}
    \lambda_i\bigl(\mathbf{n}(t^{-})-\Delta_i\bigr)
    - \lambda_i\big(\mathbf{n}(t^{-})\big) \bigg]\,\mathrm{d}t\\
\: & \quad   -\sum_i
\bigg[\ln\frac{\psi\bigl(\mathbf{n}(t^{-})+\Delta_i,t\bigr)}
  {\psi\bigl(\mathbf{n}(t^{-}),t\bigr)}
  +\beta \sum_j E_j \Delta_i^j
  + \beta\sum_{\alpha} \mu_{\alpha}\,\Delta_i^\alpha
\bigg]\,\mathrm{d}N_i(t).
\end{aligned}
\label{eq:gen_prod_expansion}
\end{equation}
Recall that in Eq.~\eqref{eq:gen_prod_expansion}, $\Delta_i^j$ and
$\Delta_i^\alpha$ are the stoichiometric changes from reaction $i$, and
$\mathbf{n}(t^{-})$ denotes the system state immediately preceding a
reaction event at time $t$.

\section{Results}

\subsection{Martingale structure of entropy production in
single-molecule chemical reaction networks}

We now evaluate the differential $\mathrm{d}(e^{-\theta(t)})$, where
$\theta(t)$ is the generalized entropy production defined in
Eq.~\eqref{eq:gen_prod}.
%
%
Using It\^{o}'s formula for semimartingales, we have

\begin{equation}
\label{eq:exp_theta}
\begin{aligned}
\mathrm{d}e^{-\theta(t)} = & -e^{-\theta(t^{-})}\sum_i \bigg[
  \frac{\psi\bigl(\mathbf{n}(t^{-})
    -\Delta_i,t\bigr)}{\psi\bigl(\mathbf{n}(t^{-}),t\bigr)}
  \lambda_i\bigl(\mathbf{n}(t^{-})-\Delta_{i}\bigr)
  -\lambda_i\bigl(\mathbf{n}(t^{-})\bigr)\bigg]\,\mathrm{d}t\\
\: & \qquad + \sum_i \bigg[e^{-\theta(t^{-}) -
  \frac{\mathrm{d}\theta(t^{-})}{\mathrm{d}N_{i}}} -
e^{-\theta(t^{-})} \bigg]\,\mathrm{d}N_i(t),
\end{aligned}
\end{equation}
where
\begin{equation}
  \label{eq:jump_in_theta}
  -\frac{\mathrm{d}\theta(t^{-})}{\mathrm{d}N_{i}}
  = \ln\frac{\psi\bigl(\mathbf{n}(t^{-})+\Delta_i,t\bigr)}{\psi\bigl(\mathbf{n}(t^{-}),t\bigr)}+
  \beta \sum_j E_j \Delta_i^j
  + \beta\sum_{\alpha} \mu_{\alpha}\,\Delta_i^\alpha.
\end{equation}
Upon substituting Eq.~\eqref{eq:jump_in_theta} into
Eq.~\eqref{eq:exp_theta}, we have

\begin{equation}
\begin{aligned}
\mathrm{d}e^{-\theta(t)} 
& = e^{-\theta(t^{-})} \biggl[
- \sum_i \Big(\frac{\psi\bigl(\mathbf{n}(t^{-})-\Delta_i,t\bigr)}
{\psi\bigl(\mathbf{n}(t^{-}),t\bigr)}
\lambda_i\bigl(\mathbf{n}(t^{-})-\Delta_i\bigr)
 - \lambda_i\big(\mathbf{n}(t^{-})\big) \Big)\,\mathrm{d}t\\
\:  & \qquad\qquad\quad  + \sum_i\Big(
\frac{\psi(\mathbf{n}(t^{-})+\Delta_i,t)}{\psi(\mathbf{n}(t^{-}),t)} e^{ \beta
\sum\limits_j E_j \Delta_i^j + \beta \sum\limits_\alpha \mu_\alpha \Delta_i^\alpha}-1
\Big)\,\mathrm{d}N_i(t)\biggr].
\end{aligned}
\end{equation}
Recall from the definition of the compensated process that
$\mathrm{d}N_i(t) = \lambda_i\bigl(\mathbf{n}(t)\bigr)\,\mathrm{d}t +
d\tilde{N}_i(t)$ is decomposed into deterministic drift and stochastic
parts. For each reaction $i$, the deterministic terms proportional to
$\dd t$ becomes

\begin{equation}
\begin{aligned}
\lambda_i\bigl( & \mathbf{n}(t^{-})\bigr)\, \biggl[\Big(1-
  \frac{\lambda_i\big(\mathbf{n}(t^{-})-\Delta_i\big)}
       {\lambda_i\big(\mathbf{n}(t^{-})\big)}
       \frac{\psi\bigl(\mathbf{n}(t^{-})-\Delta_i,t\bigr)}
            {\psi\bigl(\mathbf{n}(t^{-}),t\bigr)}
\Big) \\
\: & \qquad \qquad +
\Big(\frac{\psi\bigl(\mathbf{n}(t^{-})+\Delta_i,t\bigr)}
    {\psi\bigl(\mathbf{n}(t^{-}),t\bigr)} e^{ \beta
\sum_j E_j \Delta_i^j + \beta \sum_\alpha \mu_\alpha \Delta_i^\alpha}-1
\Big)\biggr]\\
=\: &
\frac{\psi\bigl(\mathbf{n}(t^{-})+\Delta_i,t\bigr)}
     {\psi\bigl(\mathbf{n}(t^{-}),t\bigr)}
\lambda_i\bigl(\mathbf{n}(t^{-})\bigr)e^{ \beta
\sum_j E_j \Delta_i^j + \beta \sum_\alpha \mu_\alpha \Delta_i^\alpha} 
\\
\: & \qquad \qquad -
\frac{\psi\bigl(\mathbf{n}(t^{-})-\Delta_i,t\bigr)}{\psi\bigl(\mathbf{n}(t^{-}),t\bigr)}
\lambda_i\bigl(\mathbf{n}(t^{-})-\Delta_i\bigr)
\end{aligned}
\label{eq:dt_term_simplification_i}
\end{equation}
The total deterministic drift coefficient in
$\mathrm{d}(e^{-\theta(t)})/e^{-\theta(t^{-})}$ is the sum over all pairs
of forward ($i$) and reverse ($-i$) reactions of the terms derived in
Eq.~\eqref{eq:dt_term_simplification_i}. For each such pair, this sum
is

\begin{equation}
\begin{aligned}
& \frac{\psi\bigl(\mathbf{n}(t^{-})+\Delta_i,t\bigr)}{\psi\bigl(\mathbf{n}(t^{-}),t\bigr)}
\left[\lambda_i\bigl(\mathbf{n}(t^{-})\bigr)e^{ \beta
\sum_j E_j \Delta_i^j + \beta \sum_\alpha \mu_\alpha \Delta_i^\alpha} -
\lambda_{-i}\bigl(\mathbf{n}(t^{-})+ \Delta_i\bigr)\right] \\
& \quad+
\frac{\psi\bigl(\mathbf{n}(t^{-})-\Delta_i,t\bigr)}{\psi\bigl(\mathbf{n}(t^{-}),t\bigr)}
\left[-\lambda_i\bigl(\mathbf{n}(t^{-})-\Delta_i\bigr) +
  \lambda_{-i}\bigl(\mathbf{n}(t^{-})\bigr)
  e^{-\beta \sum_j E_j \Delta_i^j - \beta \sum_\alpha \mu_\alpha \Delta_i^\alpha}\right].
\end{aligned}
\label{eq:intermediate_term}
\end{equation}

For the total drift term to vanish, it is sufficient for

\begin{equation}
e^{\beta\sum_j E_j\Delta_i^j + \beta\sum_\alpha
\mu_\alpha\Delta_i^\alpha} \lambda_i\bigl(\mathbf{n}\bigr) =
\lambda_{-i}\bigl(\mathbf{n} +\Delta_i\bigr),
\quad \forall \mathbf{n} \text{~such that } \lambda_i(\mathbf{n}) > 0
\label{eq:lambda_condition}
\end{equation}
to hold for every pair of forward ($i$) and reverse ($-i$) reactions,
ensuring that the two bracketed terms in
Eq.~\eqref{eq:intermediate_term} vanish.  In our single-molecule
setting, this condition is satisfied due to the specific form of the
propensity

\begin{equation}
\begin{aligned}
  \frac{\lambda_{-i}\bigl(\mathbf{n}(t^{-})
    +\Delta_i\bigr)}{\lambda_i\bigl(\mathbf{n}(t^{-})\bigr)} & = 
\frac{A_{-i}\,\exp\Big[-\beta\Big( G^\ddagger_{-i}
    - \!\!\!\sum\limits_{j:\,\Delta_{-i}^j<0} \!\!E_j\big(-{\Delta_{-i}^j}\big)
- \!\!\!\sum\limits_{\alpha:\,\Delta_{-i}^\alpha<0}
  \!\!\mu_{\alpha}\big(-{\Delta_{-i}^\alpha}\big) \Big) \Big] }
     {A_{i}\,\exp\Big[-\beta\Big( G^\ddagger_i - \!\!\!\sum\limits_{j:\,\Delta_i^j<0}
         \!\!E_j\big(-{\Delta_i^j}\big)
- \!\!\!\sum\limits_{\alpha:\,\Delta_i^\alpha<0}
  \!\!\mu_{\alpha}\big(-{\Delta_i^\alpha}\big) \Big) \Big] }\\
  \: & = \exp\Big[\beta \sum_{j} E_j \Delta_i^j
    + \beta \sum_{\alpha} \mu_\alpha \Delta_i^\alpha \Big]
\end{aligned}
\end{equation}
This equality holds provided that $\lambda_i(\mathbf{n}) \neq 0$,
which means the product of Kronecker deltas $\prod_{j, \Delta_i^j < 0
} \delta_{n_j}^{1}=1$ (all stochastic reactants for reaction $i$ are
present in state $\mathbf{n}$). In this single-molecule framework, if
reaction $i$ can occur to generate $\mathbf{n}+\Delta_i$ from
$\mathbf{n}$, then the reverse reaction $-i$ must be able to produce
$\mathbf{n}$ from $\mathbf{n}+\Delta_i$. Thus, the corresponding
product of Kronecker deltas for
$\lambda_{-i}\left(\mathbf{n}+\Delta_i\right)$ will also be 1 . If
$\lambda_i(\mathbf{n})=0$, then Eq.~\eqref{eq:lambda_condition}
becomes $0=0$, which is trivially satisfied since
$\lambda_{-i}\left(\mathbf{n}+\Delta_i\right)$ is also zero if
$\mathbf{n}+\Delta_i$ is an unreachable state or the reactants
necessary for $-i$ are not available.

Given that Eq.~\eqref{eq:lambda_condition} holds for single-molecule
CRNs, the entire deterministic drift component of
$\mathrm{d}e^{-\theta(t)}$ (sum of terms in
Eq.~\eqref{eq:intermediate_term}) vanishes. Consequently, only the
stochastic part, driven by the compensated Poisson processes
$\mathrm{d}\tilde{N}_i(t)$ remains and

\begin{equation}
\mathrm{d}e^{-\theta(t)} =  e^{-\theta(t^{-})} \sum_i\left(
\frac{\psi(\mathbf{n}(t^{-})+\Delta_i,t)}{\psi(\mathbf{n}(t^{-}),t)} e^{ \beta
\sum_j E_j \Delta_i^j + \beta \sum_\alpha \mu_\alpha \Delta_i^\alpha}-1
\right)\,\mathrm{d}\widetilde{N}_i(t).
\label{eq:generalized_jarzynski_crn_derivative}
\end{equation}
Eq.~\eqref{eq:generalized_jarzynski_crn_derivative} shows that
$e^{-\theta(t)}$ is a local martingale for chemical reaction networks
in which there is at most one molecule in each species.  Additional
regularity conditions analogous to the Novikov condition are needed to
ensure that $e^{-\theta(t)}$ is a martingale. Such regularity
conditions guarantee that the trajectories of $e^{- \theta(t)}$ remain
bounded.  Here, we present a relatively general regularity condition
for $e^{-\theta(t)}$ to be a martingale and leave the proof and
further discussion to Appendix~\ref{app:martingale_condition}.

\paragraph{Sufficient condition for a martingale.}
Assume that $\psi(x,t) > \varepsilon$ for all $x$ and $t$, where
$\varepsilon$ is a positive constant, then $e^{-\theta(t)}$ is a
martingale. This establishes the generalized work theorem for
single-molecule chemical reaction networks. As a result, we have for
any bounded stopping time $\tau$, conditioned on the $\sigma$-algebra
generated by $\mathbf{n}_0 \equiv \mathbf{n}(0)$,
\begin{equation}
  \mathbb{E} \big[e^{-\theta(\tau)} \big| \mathbf{n}_0\big] = 1.
  \label{eq:generalized_jarzynski_crn}
\end{equation}

\paragraph{Derivation of the standard work theorem as a special case.}
In order to obtain the original work theorem, we follow steps
analogous to those used for Langevin dynamics of continuous
variables. For a backward process, and a fixed terminal time $t_1 >
t_0=0$, we define an initial condition $\psi(\mathbf{n},t_1) =
p(\mathbf{n},t_1)$, where $p(\mathbf{n},t)$ is the forward
probability. Denote the normalized thermodynamic entropy production by
$\sigma(t) = \ln p(\mathbf{n}_0,0) - \ln p(\mathbf{n}_t, t) + \beta
Q(t)$ and introduce Manzano's correction \cite{Manzano2021February} to
$\sigma(t)$: $\sigma_{\rm M}(t) = \sigma(t) + \ln
\frac{p(\mathbf{n}_t,t)}{\psi(\mathbf{n}_t,t)} = \ln
\frac{p(\mathbf{n}_0,0)}{\psi(\mathbf{n}_t,t)} + \beta Q(t)$.  Note
that $\theta(\mathbf{n},t) = \ln
\frac{\psi(\mathbf{n}_0,0)}{\psi(\mathbf{n}_t,t)} + \beta Q(t)$ and

\begin{equation}
\begin{aligned}
  \mathbb{E}\big[e^{- \sigma_{\rm M}(t_1)} \big] &
  = \mathbb{E}\Big[e^{- \theta(t_1)}
    e^{\ln \frac{\psi(\mathbf{n}_0,0)}{p(\mathbf{n}_0,0)}} \Big]\\
  \: & = \int \Big\{\mathbb{E}\big[e^{- \theta(t_1)} \mid \mathbf{n}_0\big]
  \frac{\psi(\mathbf{n}_0,0)}{p(\mathbf{n}_0,0)} 
\Big\} p(\mathbf{n}_0,0)\mathrm{d}\mathbf{n}_0\\
\: & = \int \psi(\mathbf{n}_0,0)\mathrm{d}\mathbf{n}_0
= \int p(\mathbf{n},t_1)\mathrm{d}\mathbf{n}= 1.
\end{aligned}
\end{equation}
Since $\psi(\mathbf{n}(t_1),t_1) = p(\mathbf{n}(t_1),t_1)$ implies
$\sigma_{\rm M}(t_1) = \sigma(t_1) + \ln
\frac{p(\mathbf{n}(t_1),t_1)}{p(\mathbf{n}(t_1),t_1)} = \sigma(t_1)$,
we recover the original work theorem, otherwise known as the integral
fluctuation theorem for the total entropy production for chemical
reaction networks \cite{Schmiedl2007January}

\begin{equation}
  \mathbb{E}\big[e^{- \sigma_{\rm M}(t_1)} \big]
  = \mathbb{E}\big[e^{-\sigma(t_1)} \big] = 1.
\label{eq:jarzynski_crn}
\end{equation}

\subsection{Generalized work theorem for multi-molecule chemical reaction networks}

For general chemical reaction networks in which the number of
molecules of stochastic species is not restricted to 0 or 1,
Eq.~\eqref{eq:lambda_condition} does not necessarily hold. To see
this, we first provide the explicit expression for the propensity ratio
$\frac{\lambda_{-i}(\mathbf{n}(t^{-})+\Delta_i)}{\lambda_i(\mathbf{n}(t^{-}))}$
in the general, multi-molecule case, assuming that both
$\lambda_i(\mathbf{n}(t^{-}))$ and $\lambda_{-i}(\mathbf{n}(t^{-})
+\Delta_i)$ are non-zero,
\begin{equation}
\begin{aligned}
  \frac{\lambda_{-i}(\mathbf{n}(t^{-})+\Delta_i)}{\lambda_i(\mathbf{n}(t^{-}))} &
  = e^{\beta Q_{i}}\, \frac{\prod\limits_{j:\,\Delta_i^j>0}
    \frac{(n_j+\Delta_i)_j!}{n_j!}}{\prod\limits_{j:\,\Delta_i^j<0}
    \frac{n_j!}{(n_j+\Delta_i)_j!}},\\
  \: & = e^{\beta Q_{i}}\prod_{j} \frac{(n_j+\Delta_i)_j!}{n_j!},\\
  \: & = e^{\beta
    \big(\sum_{j} E_j \Delta_i^j + \sum_{\alpha} \mu_\alpha \Delta_i^\alpha\big)}
  \prod_j \frac{(n_j+\Delta_i)_j!}{n_j!}.
\end{aligned}
\label{eq:propensity_ratio_general}
\end{equation}
Here, $Q_{i} = \sum_{j:\,\Delta_i^j>0} \!E_j\,\Delta_i^j +
\sum_{\alpha:\,\Delta_i^\alpha>0}\! \mu_\alpha\,\Delta_i^\alpha$ is
the heat dissipation to the reservoir per reaction $i$. It is clear
that in general, $\prod_j \frac{(n_j+\Delta_i)_j!}{n_j!} \neq 1$ and
the ratio of propensities will not satisfy the form in
Eq.~\ref{eq:lambda_condition}. To remedy this, we define a new ``free
energy'' $\tilde{E}(\mathbf{n})$ by

\begin{equation}
  \tilde{E}(\mathbf{n}) \coloneqq
  \sum_j E_j\,n_j +\mfrac{1}{\beta} \sum_j \ln(n_j!).
\end{equation}
This modified energy includes the permutation entropy associated with
multiple indistinguishable molecules and ensures that entropy
production includes both energetic and state-counting contributions
\cite{Gillespie1992}. Using $\tilde{E}(\mathbf{n})$, we immediately
observe that

\begin{equation}
  \frac{\psi(\mathbf{n}(t^{-})+\Delta_i,t)}{\psi(\mathbf{n}(t^{-}),t)}
  = e^{\beta \big(\tilde{E}(\mathbf{n}(t^{-})+\Delta_i)
    - \tilde{E}(\mathbf{n}(t^{-}))
    + \sum_{\alpha} \mu_\alpha \Delta_i^\alpha\big)}.
\end{equation}

Then, we define $\tilde{\theta}$ through the differential equation
\begin{equation}
  \begin{aligned}
    \mathrm{d} \tilde{\theta}(t) & = \mathrm{d} \Sigma(t)
    + \beta \tilde{q}(t),\\
    \tilde{\theta}(0) & = 0,\\
    \tilde{q}(t) & \coloneqq - \sum_{i} \Big[\tilde{E}(\mathbf{n}+\Delta_i)
      - \tilde{E}(\mathbf{n})
      + \sum_{\alpha} \mu_\alpha \Delta_i^\alpha\Big]\,\mathrm{d}N_i(t).
  \end{aligned}
\end{equation}
By using these relationships, the previous steps from
Eq.~\eqref{eq:gen_prod_expansion} through
Eq.~\eqref{eq:lambda_condition} can be repeated, revealing the
condition
\begin{equation}
  e^{\beta \big(\tilde{E}(\mathbf{n}(t^{-})+\Delta_i)
    - \tilde{E}(\mathbf{n}(t^{-})) + \sum_{\alpha} \mu_\alpha \Delta_i^\alpha\big)}
  \lambda_i(\mathbf{n}(t^{-})) = \lambda_{-i}(\mathbf{n}(t^{-})+\Delta_i).
\label{eq:lambda_condition_general}
\end{equation}
which is straightforwardly satisfied by
Eq.~\eqref{eq:propensity_ratio_general}. As a result, the drift of
$\mathrm{d}e^{-\tilde{\theta}(t)}$ vanishes, defining
$e^{-\tilde{\theta}(t)}$ as a local martingale. Under mild conditions
such as boundedness of the propensity functions and $\psi(x,t) >
\varepsilon > 0$ for all $x$ and $t$ described in
Appendix~\ref{app:martingale_condition}, we have that
$e^{-\tilde{\theta}(t)}$ is a martingale. The additional accounting
for the multiplicity of states in $\tilde{E}(\mathbf{n})$ makes our
results different from that in Schmiedl and Seifert
\cite{Schmiedl2007January}. Hence, we have for any bounded stopping
time $\tau$,
\begin{equation}
  \mathbb{E} \big[e^{-\tilde{\theta}(\tau)} \big| \mathbf{n}_0\big] = 1,
  \label{eq:generalized_jarzynski_crn_v2}
\end{equation}
which represents a generalized work theorem for chemical reaction
networks.

\section{Numerical example}
Our generalized fluctuation theorem
(Eqs.~\ref{eq:generalized_jarzynski_crn} and
\ref{eq:generalized_jarzynski_crn_v2}) was developed to provide deeper
insight into fundamental aspects of processes in kinetic proofreading
\cite{Hopfield1974oct}, such as the trade-off between energy
consumption, operational speed, and accuracy. Our primary aim is to
numerically evaluate entropy production within a specific KPR model to
demonstrate the validity and applicability of our generalized reaction
work theorem. The entropy production of more general reactions
networks associated with biomolecular machines can also be evaluated
in this context \cite{Piephoff_hidden_2025}.

\begin{figure}[htbp]
  \centering
  \includegraphics[width=2.2in]{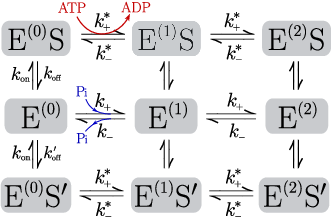}
  \caption{Schematic of a two-step kinetic proofreading network. The
    enzyme (E) can bind ``correct'' substrate S or ``incorrect''
    S$^\prime$ with equal rates $k_{\rm on}$. Correct and incorrect
    enzyme-substrate complexes (e.g., ES and ES$'$) dissociate with
    rates $k_{\rm off}$ and $k_{\rm off}'$, respectively. Bound
    complexes can undergo two successive ATP-dependent activation
    (\textit{e.g.}, phosphorylation) steps, each with rate
    $k_+^*$. Reverse steps (\textit{e.g.}, dephosphorylation) occur at
    rate $k_-^*$. Free enzyme (including following release of
    activated substrate) can also undergo
    phosphorylation/dephosphorylation with rates $k_+$ and $k_-$,
    respectively.  Activation (phosphorylation) levels of all species
    are indicated by the superscript.}
  \label{fig:two_step_detailed_KPR}
\end{figure}

Kinetic proofreading is typically described by energy-consuming
processes that enhance the specificity of enzyme (E) and substrate (S)
binding against incorrect but similar substrates (S$^\prime$).  We
choose a simple KPR reaction network that contains two activation
steps, as illustrated in Fig.~\ref{fig:two_step_detailed_KPR}.  The
chemical reaction network is framed as a random walk on a network of
enzyme states denoted by $x(t) \in
\{\mathrm{E}^{(i)},\mathrm{E}^{(i)}\mathrm{S},
\mathrm{E}^{(i)}\mathrm{S}'\}, \, i=0,1,2$. Here, correct and
incorrect substrates are assumed to have equal enzyme binding rates
$k_{\rm on}$, but slightly different unbinding rates $k_{\rm off}$ and
$k_{\rm off}^\prime$, respectively.
\begin{figure}[H]
  \centering
  \includegraphics[width=4.4in]{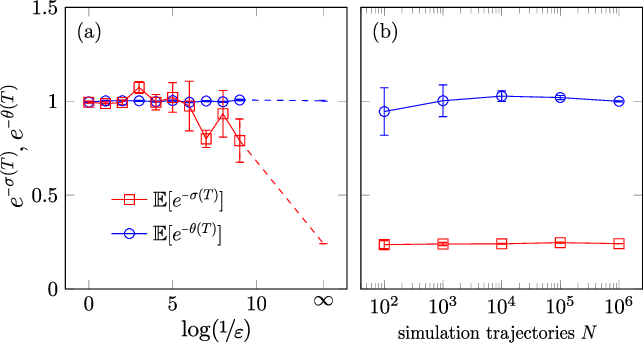}
  \caption{Numerical simulation of the single-enzyme two-step KPR
    model to evaluate $e^{-\sigma(T)}$ and $e^{-\theta(T)}$ up to a
    fixed duration $T=1$.  (a) Mean and standard error of
    $e^{-\sigma(T)}$ (the standard Jarzynski form, red squares) and
    $e^{-\theta(T)}$ (from our chemical reaction work theorem, blue
    circles) versus $\log(1/\varepsilon)$. The initial distribution is
    $p_0(x) \propto \varepsilon + \delta_{x,\text{E}}$, approaching a
    singular $\mathrm{E}^{(0)}$ initial condition as $\varepsilon \to
    0$ (i.e., $\log(1/\varepsilon) \to \infty$). Statistics are taken
    over $N=10^6$ samples.  For the generalized theorem, $\psi(x,t)$
    was set to the NESS distribution $p^*(x)$. (b) The same expected
    values and standard errors as a function of sample size $N$, for
    the strictly singular initial condition $p_0(x) =
    \delta_{x,\text{E}}$ (corresponding to $\varepsilon=0$). These
    results show the deviation from the Jarzynski equality
    (Eq.~\ref{eq:jarzynski_crn}) and validate our generalized work
    theorem (Eq.~\ref{eq:generalized_jarzynski_crn}). The parameter
    values used are $\beta=A_0=1$, $\mu_{\text{ATP}}=3.0$,
    $\mu_{\text{ADP}}=\mu_{\text{Pi}}=\mu_{\text{S}}=\mu_{\text{S}'}=0$,
    $E_{\text{E}}=0$, $E_{\text{ES}}=1.5$, $E_{\text{ES}'}=1.9$, and
    $\Delta E=1$ is the energy increase associated with each
    phosphorylation, \textit{e.g.}, $E_{\mathrm{E}^{(k)}\mathrm{S}} -
    E_{\mathrm{E}^{(k-1)}\mathrm{S}} = \Delta E$. These energy values
    correspond to $k_{+}/k_{-} = e^{-1}<1$ and
    $k_{+}^{*}/k_{-}^{*}=e^{2} >1$.}
  \label{fig:crn_fig1}
\end{figure}
Unbound enzymes can only undergo spontaneous phosphorylation and
dephosphorylation with rate $k_{+}$ and $k_{-}$, respectively. We
assume unbound enzyme phosphorylation/dephosphorylation is mediated by
inorganic phosphate (Pi) through its chemical potential $\mu_{\rm
  Pi}$.  Similarly, phosphorylation and dephosphorylation of
substrate-bound enzymes occur at rates $k_{+}^{*}$ and $k_{-}^{*}$,
respectively, and are mediated by $\mu_{\mathrm{ATP}}$ and
$\mu_{\mathrm{ADP}}$.

We first use the Gillespie algorithm \cite{Gillespie1977dec} to
numerically simulate the exponentials of the entropy production up to
a fixed final time $T=1$ and compare their expected values with the
original integral fluctuation theorem $\mathbb{E}[e^{-{\sigma}}]=1$
given by Eq.~\eqref{eq:jarzynski_crn} \cite{Schmiedl2007January}, and
with our generalized work theorem given in
Eq.~\eqref{eq:generalized_jarzynski_crn}.  For our generalized work
theorem, we choose a special backward process $\psi(x,t) = p^*(x)$,
where $p^*(x)$ is the steady-state probability density which is
readily available by numerical evaluation. Note that $p^*$ differs
from the Gibbs distribution $p_{\rm G}(x) \propto e^{-\beta E(x)}$
since coupling to ATP hydrolysis breaks detailed balance.

An interesting numerical observation arises when evaluating the
$e^{-\sigma}$ using a singular initial condition, such as $p(x,0) =
\delta_{x, \text{E}}$, where $\delta_{x, \textrm{E}}$ is the Kronecker
delta and equals 1 when $x={\rm E}$ and 0 otherwise.  To investigate
this, we regularize the initial condition using $p_0(x) \propto
\varepsilon + \delta_{x,\textrm{E}}$ and varying $\varepsilon$. As
$\varepsilon \to 0$, $\mathbb{E}[e^{-\theta}] \approx 1$ while
$\mathbb{E}[e^{-\sigma}]$ becomes unstable and drops below 1 as
$\varepsilon \to 0$, shown in Fig.~\ref{fig:crn_fig1}(a).

The deviation of $\mathbb{E}[e^{-\sigma}]$ from 1 when $\varepsilon =
0$ is persistent as the number of samples increases, shown in
Fig.~\ref{fig:crn_fig1}(b). The failure of
Eq.~\eqref{eq:jarzynski_crn} for singular initial conditions may be
related to failure of the local martingale to be a true martingale.
It is noteworthy that the sufficient condition discussed in
Appendix~\ref{app:martingale_condition}, $\psi(x,t) > \varepsilon > 0$
for all $x$ and $t$, is not satisfied when $\varepsilon = 0$.

Next, we investigate a scenario involving a stopping time. The process
is stopped at time $\tau = \min(\tau_{{\text{target}}}, T)$, where
$\tau_{{\text{target}}}$ is the first passage time to a specified
target state. The system's state $x(\tau)$ that defines the stopping
time $\tau$ can be of biological relevance.
\begin{figure}[H]
  \centering
  \includegraphics[width=2.8in]{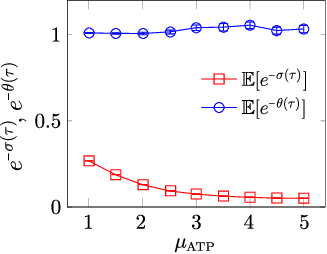}
  \caption{Starting from a singular initial condition at
    $\mathrm{E}^{(0)}$, we simulated the single-enzyme KPR process up
    to a stopping time $\tau$ defined as the first passage time to any
    one of the fully phosphorylated states ($\mathrm{E}^{(2)}$,
    $\mathrm{E}^{(2)}\mathrm{S}$, or $\mathrm{E}^{(2)}\mathrm{S}'$),
    bounded by a maximum time $T=4$. Different values of the ATP
    chemical potential $\mu_{\rm ATP}$ are used.  Evaluating
    $e^{-\sigma(\tau)}$ and $e^{-\theta(\tau)}$ over $N=10^6$
    independent samples, we plot the mean and standard errors of
    $e^{-\sigma(\tau)}$ and $e^{-\theta(\tau)}$ to show the deviation
    from the Jarzynski equality (Eq.~\ref{eq:jarzynski_crn}) and again
    validate the generalized work theorem
    (Eq.~\ref{eq:generalized_jarzynski_crn}). Parameters values are
    set to those used in Fig.~\ref{fig:crn_fig1} except that $\mu_{\rm
      ATP}$ is varied.}
    \label{fig:crn_fig2}
\end{figure}
For example, reaching $x(\tau) = \mathrm{E}^{(2)}\mathrm{S}$ (a fully
activated correct substrate) might indicate successful recognition and
processing, while other outcomes could signify errors or incomplete
processing.  For simplicity, in this numerical example, we define the
target states as any of the fully phosphorylated configurations
$\mathrm{E}^{(2)}$, $\mathrm{E}^{(2)}\mathrm{S}$, or
$\mathrm{E}^{(2)}\mathrm{S}'$ and $T=4$ is a preset maximum
duration. For these simulations, the system starts from the singular
initial condition $x_0 = \mathrm{E}^{(0)}$ (unbound, unphosphorylated
enzyme) with probability 1. Due to this singular initial condition and
the stopping time $\tau$, the standard fluctuation theorem
($\mathbb{E}[e^{-\sigma}]=1$) is not directly applicable. However, our
generalized theorem, based on the martingale property of
$e^{-\theta(t)}$ (where $\psi=p^*$), predicts that
$\mathbb{E}\big[e^{-\theta(\tau)} \mid x_0=\mathrm{E}^{(0)}\big] = 1$,
because $\tau$ is a bounded stopping time (since $\tau \le T$).  We
test this prediction numerically and show the results in
Fig.~\ref{fig:crn_fig2}.

%
\begin{figure}[htbp]
  \centering
  \includegraphics[width=2.8in]{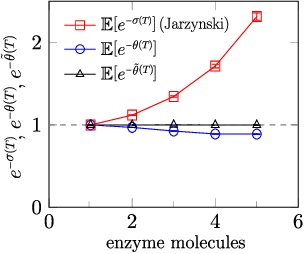}
  \caption{Numerical validation of the KPR system shown in
    Fig.~\ref{fig:two_step_detailed_KPR} when multiple enzyme
    molecules participate. In this example we set $\mu_{\rm ATP}=3$,
    use a uniform initial distribution, and evaluate the entropy
    produced up to fixed time $T=1$ over $10^6$ simulated
    realizations. This analysis was performed for systems with
    1,2,3,4, and 5 enzyme molecules.  The results validate the
    generalized work theorem in
    Eq.~\ref{eq:generalized_jarzynski_crn_v2} based on the
    multiplicity-corrected free energy $\tilde{E}$.}
  \label{fig:crn_fig3}
\end{figure}
Finally, we consider an example in which the number of enzyme
molecules exceed one. Quantities involving $\sigma$ (Jarzynski's
original formulation \cite{Jarzynski1997April}, without our backward
process), $\theta$ (without multiplicity correction), and
$\tilde{\theta}$ (with multiplicity correction in the free energy)
were simulated and are shown with their standard errors in
Fig.~\ref{fig:crn_fig3}.  We assumed a uniform initial
distribution--every distinct state has the same probability and used
$10^6$ simulation trajectories for each data point.

\section{Discussion}

By using a generalized entropy production $\theta$ that is defined
relative to an auxiliary backward process rather than forward
probability on which the conventional entropy production $\sigma$ is
based, we established a generalized work theorem for stochastic
chemical reaction networks that holds for arbitrary, bounded stopping
times. Our central contribution is the identification of
$e^{-\theta(t)}$ as a martingale in single-molecule CRNs, extending
fluctuation theorems to discrete single-molecule biochemical
systems. It is noteworthy that this result is not based on any
detailed balance assumption, since the relation in
Eq.~\eqref{eq:lambda_condition} does not depend on any probability
distribution over the states of the system. Instead, it arises from
the definition of reaction propensities in Eq.~\eqref{eq:propensity}.

In more general cases where there can be more than one molecule in the
system, Eq.~\eqref{eq:lambda_condition} fails to make $e^{-\theta(t)}$
a martingale.  Physically, this failure arises because the permutation
contribution to the energy of the stochastic species is not
considered. For chemical reaction networks involving multiple
molecules, we invoke a free energy that includes the associated
combinatorial entropy.  The entropy production $\tilde{\theta}$
associated with this modified free energy $\tilde{E}$ renders
$e^{-\tilde{\theta}(t)}$ a martingale, further generalizing our work
theorem to chemical reactions that involve an arbitrary number of
reactant molecules.




The stopping-time generality of our results makes them particularly
valuable for biological processes where termination conditions are
state-dependent rather than time-dependent. We propose that concepts
such as the energy-accuracy tradeoff in KPR-like biochemical reaction
networks \cite{Lindsay_KPR,Li_KPR,Xiao_KPR} can be analyzed through
the martingale properties of entropy production associated with
stopping times. Similarly, for epigenetic regulation involving rare
state transitions \cite{Li_histone}, the generalized work theorem
provides thermodynamic constraints that must hold regardless of when
transitions occur. These applications highlight how our results bridge
thermodynamic and information-theoretic perspectives of
decision-making in biochemical reaction networks
\cite{esposito2020open}.

Mathematically, we restricted ourselves to bounded stopping times
$\tau \leq T$, ensuring that the optional stopping theorem applies
directly to the martingale $e^{-\theta(\tau)}$ or
$e^{-\tilde{\theta}(\tau)}$, and guaranting the validity of our
identities (Eqs.\ref{eq:generalized_jarzynski_crn} and
~\ref{eq:generalized_jarzynski_crn_v2}) without further integrability
assumptions. This requirement on boundedness originates from using
Doob's optional sampling theorem \cite{Williams_1991} to obtain the
martingale identity and can be relaxed in multiple ways, particularly
when the underlying stochastic process possesses certain properties
such as boundedness. Exploring physically realistic conditions when
the martingale condition fails can enhance our general understanding
of stochastic thermodynamics.

\bibliography{work_theorem.bib}
\bibliographystyle{plain}
\appendix

\section{On the martingale property of stochastic chemical reaction
networks}
\label{app:martingale_condition}
In general, a local martingale $M_t $ is a martingale if and only if
it is of class DL, \textit{i.e.}, for any $a >0$, $\{M_{\tau}: \tau
\leq a ~\text{is a stopping time}\} $ is a uniformly integrable
family.

In the setting of Eq.~\eqref{eq:generalized_jarzynski_crn_derivative},
assume that $\psi(x,t) > \varepsilon$ for some constant $\varepsilon > 0$,
Since $\psi(x,t)$ is bounded above, we have 

\begin{equation}
  \Delta \Sigma(t) \leq 2 \ln \frac{1}{\varepsilon}.
  \label{eq:delta_sigma}
\end{equation}
As a result, $|\sigma(t)| \leq 2 \ln \frac{1}{\varepsilon} + \beta
|Q(t)|$.

In single-molecule or finite-molecule systems, ${\mathbf{n}} $ is
bounded. Consequently, the reaction rates $\lambda_i(x)$ are also
bounded.  Denote the maximum reaction rate by $\Lambda$.  Without loss
of generality, we only consider one reaction $i$, whose occurrence is
denoted by $N_i(t)$.

As is discussed previously, $N_i(t)$ is a time-changed Poisson process
with intensity $\lambda_i(x(t))$. We can find a standard Poisson
process $Y(t)$ such that

\begin{equation}
N_i(t) = Y \left( \int_0^t \lambda_i(x(s))\,\mathrm{d}s \right) \leq Y
\left( \Lambda t \right),
\end{equation}
where the inequality holds because $\lambda_i(x)$ is bounded by $\Lambda$ and 
Poisson process $Y(t)$ is non-decreasing.

Similarly, we denote $\overline{\Delta} = \max_i \left\{ \sum_j
|E_j({\mathbf{n} + \Delta_i}) - E_{\mathbf{n}} | + \sum_\alpha
|\mu_\alpha({\mathbf{n} + \Delta_i}) - \mu_\alpha({\mathbf{n}})|
\right\} $. Then, we have a bound on the entropy production by

\begin{equation}
  \beta |Q(t)| \leq \beta \overline{\Delta} Y_{\Lambda t},
  \label{eq:beta_Q}
\end{equation}
where $Y_{\Lambda t}$ is a standard Poisson process with intensity
$\Lambda t$.

Combining Eq.~\eqref{eq:delta_sigma} and Eq.~\eqref{eq:beta_Q}, we
have
\begin{equation}
  |\sigma(t)| \leq 2 \ln \mfrac{1}{\varepsilon} + \beta
  \overline{\Delta} Y_{\Lambda t}, ~ \text{a.s.}
\end{equation}
As a result, for all stopping times $\tau \leq a$, we have

\begin{equation}
  \begin{aligned}
    e^{- \sigma(\tau)} &\leq \Big( \mfrac{1}{\varepsilon} \Big)^2
    e^{ \beta \overline{\Delta} Y_{\Lambda \tau}} \leq \Big(
    \mfrac{1}{\varepsilon} \Big)^2 e^{ \beta Y_{ \overline{\Delta}
        \Lambda a}}, ~ \text{a.s.}
  \end{aligned}
\end{equation}

It is straightforward to show that $\mathbb{E}\left[e^{ \beta Y_{
      \overline{\Delta} \Lambda a}}\right] < \infty$.  As a result, we
have $\{e^{- \sigma(\tau)} : \text{stopping time}\,\, \tau \leq a \}$
is uniformly integrable:

\begin{equation}
  \sup_{\tau \leq a}\mathbb{E}\left[e^{- \sigma(\tau)}\right] < \infty.
\end{equation}
We conclude that $e^{- \sigma(t)}$ is of class DL and is thus a
martingale.

\end{document}